\documentclass[twocolumn,trackchanges]{aastex63}
\usepackage{amsmath}
\usepackage{graphicx}
\usepackage{amsfonts}
\usepackage{natbib}
\usepackage{color}
\usepackage{epstopdf}
\usepackage{graphicx}
\usepackage{multirow}
\usepackage{longtable}
\usepackage{rotating}

\shorttitle{Radio QPO in NLSy1 galaxy J0849+5108}
\shortauthors{Zhang \& Wang}

\newcommand\fermi{{\it Fermi}}
\newcommand\gr{{$\gamma$-ray}}

\begin{document}

\title{A Radio Quasi-periodic Oscillation of 176 day in the Radio-loud Narrow-line Seyfert 1 Galaxy J0849+5108}

\author{Pengfei Zhang}
\affil{Department of Astronomy, School of Physics and Astronomy, Key Laboratory of Astroparticle Physics of Yunnan Province, Yunnan University, Kunming 650091, People’s Republic of China; zhangpengfei@ynu.edu.cn; wangzx20@ynu.edu.cn}

\author{Zhongxiang Wang}
\affil{Department of Astronomy, School of Physics and Astronomy, Key Laboratory of Astroparticle Physics of Yunnan Province, Yunnan University, Kunming 650091, People’s Republic of China; zhangpengfei@ynu.edu.cn; wangzx20@ynu.edu.cn}
\affil{Shanghai Astronomical Observatory, Chinese Academy of Sciences, 80 Nandan Road, Shanghai 200030, People’s Republic of China}

\begin{abstract}

We analyze the 11-year long-term light curve of the Radio-loud Narrow-line 
Seyfert 1 (NLSy1) galaxy J0849+5108 and the nearly simultaneous \gr\ data
of the source. The data were obtained with the Owens Valley Radio 
Observatory 40-m telescope at 15~GHz and 
with the Large Area Telescope onboard {\it Fermi Gamma-ray Space Telescope},
respectively.
A quasi-periodic oscillation (QPO) signal at a significance of $>$5$\sigma$
is found in the radio light curve, but no similar modulation is seen in the 
\gr\ light curve. The QPO signal was present for 21 cycles, the longest
among the reported radio QPOs, and
likely disappeared after 2019 January 16. Different mechanisms 
proposed to explain the observed QPOs in Active Galactic Nuclei (AGN) are 
discussed for this QPO case. { Either a secular instability in the inner
accretion disk or a helical
structure in the jet of J0849+5108 may explain the radio QPO, while for
the latter scenario
the jet has to be collimated up to $\sim$200~pc, similar to that seen in the
nearby AGN M87.} It will be of interests to keep monitoring the source at radio
frequencies, seeing if similar QPO signals would appear again or not.
\end{abstract}


\section{Introduction}
\label{sec:intro}

Active Galactic Nuclei (AGN) are objects of great interest, as
they contain an active supermassive black hole (SMBH) at their center 
and $\sim$10\% of them have relativistic jets.
Emission from them ranging from MHz radio frequencies to TeV \gr\ energies
can be mostly dominated by non-thermal radiation and
variable on a variety of timescales and amplitudes.
Among phenomena related to their variability, there is very intruiging one, 
the so-called quasi-periodic oscillations (QPOs), which have more and more 
cases reported recently.
The QPOs were found at all frequencies from radio to GeV (or even TeV) 
$\gamma$-rays and their periods span over a wide range of timescales 
from minutes to years.

After the launch of {\it the Fermi Gamma-ray Space Telescope (Fermi)}, 
a few cases of periodic signals in $\gamma$-rays from AGN were reported.
Possible QPOs of PG 1553+113, PKS 2155$-$304, PKS 0301$-$243, PKS 0537$-$441, PKS 0426$-$380, PKS 0601$-$70, and PMN J0948+0022 have been studied by
\citet{Ackermann2015}, \citet{Sandrinelli2014,Sandrinelli2016a,Sandrinelli2016b}, \citet{Zhang2017a,Zhang2017b,Zhang2017c,Zhang2020}, and \citet{Zhangjin}.
Their periods span in the range of $\sim$ 1--3 years. Interestingly, in a 
blazar-type AGN PKS~2247$-$131, a 34.5-day QPO after an initial flux peak of an
outburst was reported by \citet{Zhou2018}, and it has been the only one having 
relatively short, monthly period in $\gamma$-rays thus far.

In X-rays, a significant QPO was reported in RE~J1034+396 
by \citet{Gierlinski2008}. Then other possible QPOs were detected in 
Mrk 766, 2XMM J123103.2+110648, and 1H 0707$-$495 
\citep{Zhang2017d,Lin2013,Pan2016,Zhang2018}.
The peculiar case is in 1H~0707$-$495, in which double QPOs appeared and were
separated 
by an intermediate state with a frequency ratio of $\sim$1:2 \citep{Zhang2018}.
Different from those found in $\gamma$-rays, these QPOs have very short periods,
$\sim$1--3 hrs. They possibly follow the correlation between QPO 
frequency and black-hole mass $M_{\rm BH}$, as suggested in 
\citet{Kluzniak2002,Remillard2006,Zhou2010,Zhou2015,Pan2016,Zhang2017d,Zhang2018}.

In optical band, the most well-known case is OJ 287, which exhibits $\sim$ 12-yr
periodicity in its optical light curve based on more than a century 
monitoring \citep{Kidger1992,Fan2000,Valtonen2006,Valtonen2008}.
In PKS 2155$-$304, a 317~day QPO was found in the optical light curve 
on a $\sim$3600-day long-term time scale \citep{Zhang2014}. Interestingly this 
period is exactly half of that of the QPO detected in $\gamma$-rays \citep{Sandrinelli2014,Zhang2017a}.

Several cases of AGN QPOs were also found in radio light curves. 
In CGRABS~J1359+4011, a QPO exhibiting 150 day period was reported 
by \citet{King2013}. \citet{Bhatta2017} claimed a QPO having 270-day period 
in PKS 0219$-$164. 1ES 1959+650, J1043+2408, and PKS J0805$-$0111 were also 
reported to have 
possible QPO behavior \citep{Li2017,Bhatta2018,Ren2020}. These radio QPOs 
have periods ranging from approximately half a year up to several years, 
similar to those of QPOs found in $\gamma$-rays.

{ While the drastically different timescales of the periodic signals 
found in such as X-rays and
$\gamma$-rays are likely due to the selection effect caused by the observational
modes---no sensitive X-ray telescopes would spend valuable time to monitor 
AGN for years (also see \citealt{Vaughan2005} for detectability of
X-ray QPOs in AGN) and \fermi\ does not have sufficiently high sensitivity to 
monitor many AGN on minute-long timescales, they have been discussed to
indicate different physical processes. For example, the sub-day QPOs
have been compared to those found in stellar-mass black hole systems
and are thought to similarly reflect accretion activities near a SMBH 
(e.g., \citealt{Gierlinski2008}), while the year-long ones are generally 
considered to be the indicator of binary SMBH systems 
(e.g., \citealt{Ackermann2015}). 
Because of the middle position in the periodicity range of QPOs, the 
month-long one in PKS~2247$-$131 was interpreted to be due to
the helical structure in the jet \citep{Zhou2018}. Therefore QPOs could 
serve as a unique probe to help reveal structures and physical processes 
of AGN.}

In this paper we report a discovery of a QPO at $\sim$176 day with a 
significance of over $5\sigma$ in radio emission from a 
Radio-loud Narrow-line Seyfert 1 (NLSy1) galaxy, J084957.97+510829.0 
(hereafter J0849+5108). The radio light curve was obtained from monitoring 
observations of J0849+5108 with the Owens Valley Radio Observatory (OVRO) 40-m 
telescope. As \emph{Fermi} provided monitoring of the source nearly 
simultaneous, we also analyzed the \gr\ data. Below we first 
summarize the properties of J0849+5108 in section~\ref{subsec:sum}.
In section \ref{data}, we describe the analysis for the OVRO radio light curve 
and \emph{Fermi} \gr\ data and show the main results. 
In section \ref{sumdis} the results are discussed. 

{\subsection{NLSy1 galaxy J0849+5108}
\label{subsec:sum}
The NLSy1 galaxies are a subclass of AGN \citep{Osterbrock1985,Mathur2000}, 
containing black holes at the low end of the AGN $M_{\rm BH}$ range 
\citep[$\rm 10^5-10^8~M_{\odot}$;][]{Zhou2006,Xu2012} and having
high Eddington accretion rates (between 0.1--1; \citealt{Boller1996}).
A few percent of them are radio loud, indicating the presence of relativistic 
jets \citep{Komossa2006,Lister2018}, 
and some show jet morphologies on parsec scales \citep{Doi2007,Gu2010}.
J0849+5108 is one of the most studied NLSy1 galaxies 
(see \citealt{Dammando2012,Maune2014} and references therein). The 
central black hole was estimated to have  
$M_{\rm BH}\sim 10^{7.4}~M_{\odot}$ \citep{Yuan2008}. 
Since 2011 June, \gr\ flares from it were detected with \fermi. Combined 
with radio studies, the high-energy detection confirmed the presence of a jet 
in it \citep{Dammando2012}. 
High-resolution radio observations resolved
its radio emission, and a jet structure was seen extending from the radio core
to several tens of parcsec projected distance \citep{Dammando2012,laa2018}.
Based on broadband studies and its strong brightness variability, its 
similarity with the blazar-like AGN \citep{Maune2014} has been established, 
i.e., the jet is pointing close to our line of sight \citep{Paliya2016}. 

In this paper, the cosmological parameters from the Planck mission 
\citep{Planck2014} are used (the Hubble constant 
$H_0\simeq 67$\,km\,s$^{-1}$\,Mpc$^{-1}$). The redshift of the source 
$z \simeq 0.584$ \citep{Alam2015} corresponds to the luminosity distance 
of $\sim$3500\,Mpc.
}

\begin{figure*}
\centering
\includegraphics[width=0.62\textwidth]{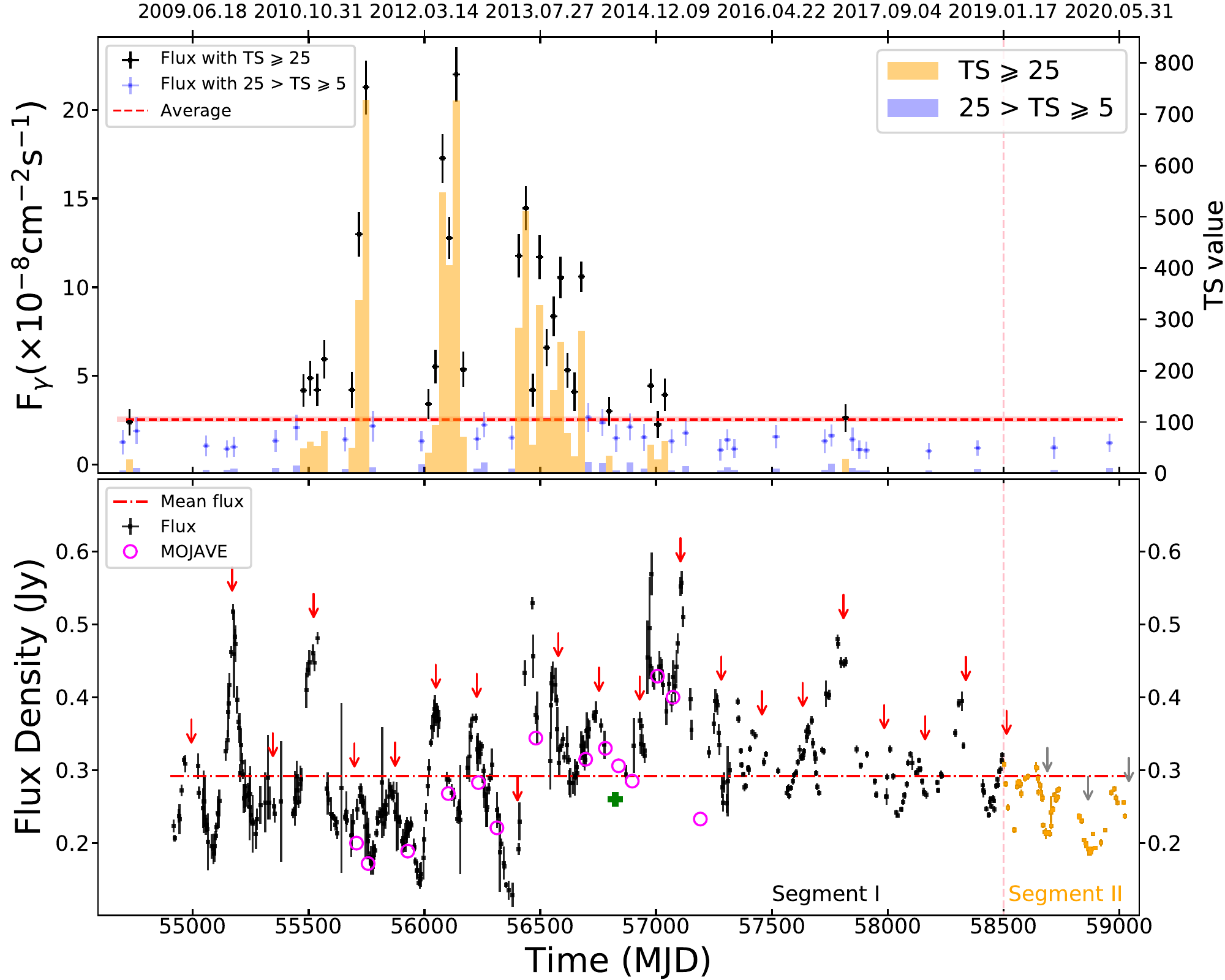}
\caption{Light curves of the NLSy1 galaxy J0849+5108. 
{\it Bottom} panel: 15-GHz 
flux density curve obtained with the OVRO 40-m telescope. 
The green cross marks the maximum interval between two data points and
the red horizontal dotted-dashed line is the mean flux density of 
the radio light curve.  The pink vertical dashed line marks the position 
after which the QPO vanishes (data points in orange color), and the arrows 
indicate the expected flux maxima for every cycles. {The peak 
intensity measurements of the source's radio emission at 15~GHz from the MOJAVE 
survey \citep{laa2018} are indicated by purple circles.}
{\it Top} panel: 0.1--500 GeV \gr\ monthly light curve, {
	for which the TS values of the data points are given by orange 
	(TS$\geqslant 25$) and light blue ($25\geqslant$TS$\geqslant 5$) bars.
	The data points in these two TS ranges are marked with black and light
	blue color correspondingly.} The red dashed line represents the average
	integrated photon flux over the whole time period.}
\label{Fig1}
\end{figure*}

\section{Data Analysis and Results}
\label{data}
\subsection{OVRO light curve data}
The OVRO 40-m telescope has been monitoring over 1800 
AGN\footnote{https://sites.astro.caltech.edu/ovroblazars/index.php?page=sourcelist} twice a week at 15~GHz since 2008.
The obtained data are reduced and calibrated to form light curves available on 
their publicly accessible database\footnote{https://sites.astro.caltech.edu/ovroblazars}.
The flux density scale is derived using the primary calibrator {3C~286}, 
and the detailed information about the observing program, data reduction, 
and calibration procedure are described in \citet{Readhead1989,Richards2011}.
The source J0849+5108 (R.A. = $08^h49^m58^s.080$, 
decl. = $51^{\circ}08^{'}29^{''}.040$) is one of the OVRO's targets. It has 
been observed for approximately 11 years (from 2009 March 28 to 2020 June 22). 
We show its radio light curve in the bottom panel of Figure~\ref{Fig1}.
The minimum and median time intervals of two adjacent time bins are $\sim$0.68 
and 5.6 days respectively, and the maximum interval is $\sim$92 days at 
MJD 56823.6 (marked with a green cross in Figure~\ref{Fig1}).
In the light curve, the minimum (at MJD 56380.32), maximum (at MJD 56980.71), 
and mean value (shown with a red horizontal dotted-dashed line) 
are 0.12, 0.57, and 0.29 Jy respectively, and its standard deviation is 7.7\%.

\subsection{{\it Fermi} Large Area Telescope data}
The source J0849+5108 was detected with the Large Area Telescope
(LAT) onboard \emph{Fermi} \citep{Atwood2009}. It is named as 
4FGL J0850.0+5108 in the fourth Fermi Large Area Telescope catalog 
\citep[4FGL;][]{Abdollahi2020}.
We selected \emph{Fermi}-LAT 0.1--500~GeV Pass 8 (\emph{Front+Back} SOURCE
class) photon-like events from the LAT data between 2008 August 4 and 2020 
August 28 within a $20^{\circ}\times20^{\circ}$ region centered at 
the position of J0849+5108. The events were reduced by requiring the zenith 
angle $\rm <90^\circ,~DATA\_QUAL > 0, and ~LAT\_CONFIG=1$.
A binned maximum likelihood analysis was performed to the selected events data 
using the instrument response files (P8R3\_SOURCE\_V2) and the 4FGL 
model \citep{Abdollahi2020}. In 4FGL, emission of J0849+5108 is described 
with a log-parabolic spectrum model.
From the analysis, the integrated photon flux in the 0.1--500~GeV energy range 
was obtained to be $\rm (22.8\pm1.3)\times10^{-9}~photons\,cm^{-2}\,s^{-1}$ 
with a test statistic (TS) value of $\sim$ 2940. The best-fit photon spectral 
parameters $\alpha$, $\beta$, and $E_b$ were $2.18\pm0.04$, $0.07\pm0.02$, 
and $649\pm90$ MeV, respectively. {These values are in agreement with those
reported in 4FGL \citep{Abdollahi2020}.}
The best-fitting results were saved as a new model file. We then obtained
light curves of the source based on this new model file.
\begin{figure*}
\centering
\includegraphics[scale=0.5]{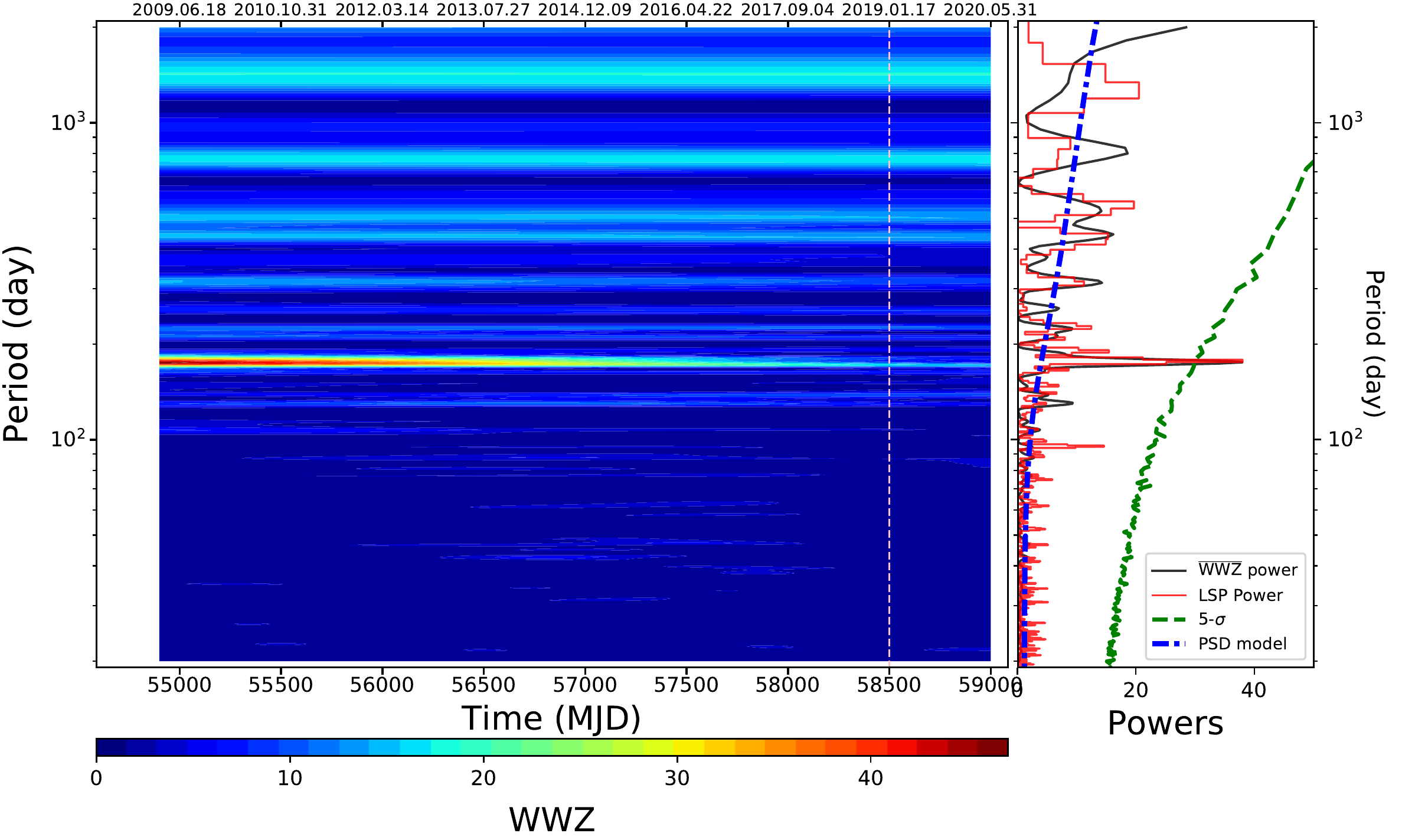}
\caption{Power spectra of the J0849+5108 OVRO light curve. {\it Left} panel: the color-scaled WWZ power spectrum calculated in a range of frequencies over the time duration of the whole light curve. 
     {\it Right} panel: time-averaged WWZ power spectrum (black curve) and LSP power spectrum (red histogram), calculated from the data before MJD~58500 (marked by the dashed vertical line in the left panel). The blue dashed-dotted line is the best-fit underlying PSD, and the green dashed line is the $5\sigma$ confidence curve evaluated from artificial light curves.}
\label{Fig2}
\end{figure*}

{The source's \gr\ emission was generally due to three major flares, 
although the third one (at MJD~56500) shows complex variation structures.
Different time bins were tested when we constructed a light curve, and 
we found that a monthly (30-day bin) light curve depicts the overall 
variation features of the source (see Figure~\ref{Fig1}).
For this light curve, we considered a data point 
with TS$\geqslant$25 ($\gtrsim 5\sigma$) as a detection, while 
those data points with $25\geqslant$TS$\geqslant$5 were also included
in order to show the flux
variations as complete as possible. The TS values of the data points are 
given in Figure~\ref{Fig1}, where the two TS ranges are marked with orange 
and light blue color respectively.}
\subsection{Temporal variability analysis and results}
{To search for periodicity in the OVRO radio light curve of J0849+5108 by 
calculating 
a power spectrum, we first employed the method of the Weighted Wavelet 
Z-transform \citep[WWZ;][]{Foster1996}.
In WWZ, the wavelet transform is casted as a weighted projection 
and observed fluxes are compared to Morlet wavelet functions instead of 
sinusoids \citep{Foster1996}, and the method captures frequency and 
location information (here in time) simultaneously by changing the parameters 
of wavelets. For non-persistent periodic signals, WWZ can show when a periodic 
signal starts to decrease. We obtained} color-scaled WWZ power spectrum for 
the whole light curve, which is shown in the left panel of
Fig.~\ref{Fig2}. A signal is seen at $\sim 176$~days. The time-averaged WWZ 
power spectrum (averaged at frequencies along time) was also obtained,
and careful examination of the power value of it allowed us to find that 
the power value increases to the maximum at MJD 58499.22 (2019 January 16, 
marked with pink vertical dashed line in Figure~\ref{Fig1}) but then
starts decreasing.
We thus divided the light curve into two segments (see the bottom panel 
of Figure~\ref{Fig1}): Segment I (black) and Segment II (orange).
The time-averaged WWZ power spectrum in Segment I is shown in the right panel 
of Figure~\ref{Fig2}. A QPO signal has a peak at $175.93\pm6.34$ days, where
the uncertainty was estimated as the full width at half maximum of a 
Gaussian function that fits the power peak. 

We then used the generalized Lomb-Scargle 
Peridogram \citep[LSP;][]{Lomb1976,Scarle1982,Zechmeister2009}  
as an independent check for the result from the WWZ method. 
{The LSP method is a common tool in time series analysis for unequally 
spaced 
data, and can provide accurate frequencies and power spectral 
intensities \citep{Zechmeister2009}.  Only the light curve in Segment~I was 
analyzed. The resulting power spectrum is shown as the red histogram in 
Figure~\ref{Fig2}. A sharp peak around 176 days, nearly the same as that
from the WWZ method, was revealed. The results confirm the signal detection
from the WWZ method.}

We used the period to calculate the expected flux maxima for every cycles, 
which are shown as red arrows in the bottom panel of Figure~\ref{Fig1}. 
It can be seen that although the light curve overall has large variations, 
the arrows in most cases point at the local flux maxima in the light curve. 
It can also be noted that in Segment II the arrows start missing the flux 
maxima, supporting the result from our above WWZ analysis that the signal 
starts decreasing after $\sim$MJD~58500.

For the power peak, the probability $p$ to obtain a power equal to or higher 
than the threshold from a chance fluctuation is $< 5.9\times10^{-19}$, 
and the false-alarm probability (FAP $= 1 - (1 - p)^N$) is 
$1.2\times10^{-16}$ \citep{Horne1986,Zechmeister2009}, where $N$ is the number 
of independent frequencies sampled (i.e., the trial factor).
To estimate the underlying power spectral density (PSD), we used a function
of smoothly bending power-law plus a constant \citep{Gonzalez2012} to model 
the PSD calculated from the light curve. A maximum likelihood method was 
employed \citep{Barret2012}.
The PSD function has a form of $P(f)=Af^{-\alpha}[1+(f/f_{bend})^{\beta - \alpha}]^{-1} +C$, where $A$, $\alpha$, $\beta$, $f_{bend}$, and $C$ are the normalization, low frequency slope,
high frequency slope, bend frequency, and Poisson noise, respectively, and 
the obtained values are $1.12\pm0.84$, $0.33\pm0.03$, $2.73\pm0.69$, $0.0059\pm0.0016$, and $1.19\pm0.06$, respectively.
To evaluate a confidence level for the periodic signal, we generated $10^7$ 
artificial light curves based on the best-fit underlying PSD and probability 
density function of the flux density variations by employing the simulation 
program provided in \citet{Emmanoulopoulos2013}.
The confidence curve evaluated from the artificial light curves is shown as 
a green line in the right panel of Figure~\ref{Fig2}. The confidence level 
for the QPO signal was found to be $> 5\sigma$.

We also checked the \gr\ light curve for any similar QPO signals.
However the light curve mainly consists of three major flares and no QPO 
signals were found. We note that the third flare shows repeated sub-flare
structures, which is discussed below.

\section{Discussion}
\label{sumdis}

We have analyzed the OVRO 15-GHz light curve of J0849+5108
from 2009 March to 2020 June, and found a QPO with a period of 176 day
at a significance of $>5\sigma$ in the data before 2019 January.
Compared to the previous QPO cases reported in radio light curves of AGN, 
this QPO lasted the most cycles (21 cycles) and had a significance similar to 
or even higher than most of the others.
The QPO signal likely disappeared after 2019 January; 
for example, if the whole light curve is considered,
the significance of the signal would become much lower. 
This transient nature makes it fit the other
reported QPOs, as most of them are transient phenomena.
It is of interest to keep monitoring the source, checking how its radio
emission varies and whether or not the QPO signal would appear again.
\begin{figure}
\centering
\includegraphics[scale=0.8]{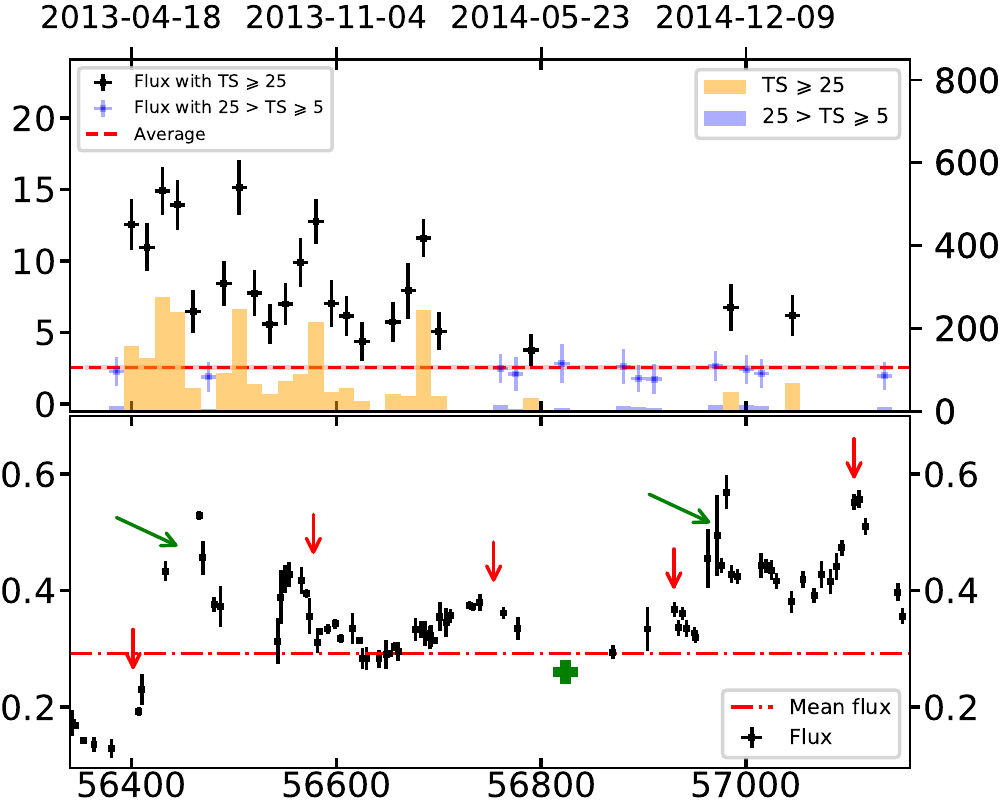}
	\caption{Same as Figure~\ref{Fig1}, but the time bin of the \gr\
	light curve is 15 days. Two radio flux density peaks, in addition to
	the peaks of the QPO cycles, are marked by green arrows. }
\label{fig3}
\end{figure}

The nearly simultaneous {\it Fermi}-LAT data were also analyzed, but only 
three major flaring events were detected and no similar periodic pattern 
was seen. \citet{Dammando2013} suggested that the \gr\ flare around 
MJD~56100 was 
connected to the radio brightening around MJD~56200--56300 (i.e., related but 
delayed radio activity), but our result indicates that the radio
brightening actually was only one of the QPO cycles, not likely 
related to the \gr\ flaring events. {However the delayed radio
activity maybe was seen in the third major flare in year 2013. 
\citet{Maune2014} presented
broadband monitoring data covering the early half of year 2013, which
contain the OVRO light curve during the time period. In Figure~\ref{fig3},
we show the details of the \gr\ and radio light curves in 2013--2014, 
where we used 15 day for the time bin of the former. The 2013 flare appears to 
consist of four subflares (indicated by fluxes and $>200$ TS values), 
the first of which was discussed 
in \citet{Maune2014}. They reported optical and X-ray flux as well as optical 
polarization measurements, which show significant variations related to 
the \gr\ flare (or the first subflare), and suggested a radio peak 
(marked by the first green arrow in Figure~\ref{fig3}) as the delayed activity. This radio peak
indeed is not one of the QPO cycles, and also we note that the following peak 
mismatches (and occurs before) the red arrow by $\sim$25 days 
(four times that the uncertainty of the QPO period). In addition, the radio peak
before MJD~57000 (marked by the second green arrow) is similarly not
one of the QPO cycles, and it coincides with 
a minor \gr\ flaring event---a TS$>200$ \gr\ 
data point around the same time. Therefore it is reasonable to consider
that these radio variations were possibly related to the \gr\ flaring 
activities.}
{While} it has been established from simultaneous monitoring at radio and 
$\gamma$-rays that few of blazars show correlated variations at the two 
bands \citep{Max-Moerbeck2014}, {we suggest that the radio emission of
J0849+5108 may contain two components: one the baseline that shows the
QPO and the other related to additional activities of the system including
the \gr\ flares.}

Possible mechanisms behind the QPOs in AGN have been widely discussed 
(see, e.g., \citealt{King2013,Ackermann2015} and references therein). 
Since QPOs in stellar-mass black-hole 
binary systems have been well studied, and in most cases they were interpreted 
to be related to 
the accretion in the innermost stable circular orbit around
black holes \citep{Remillard2006}, the similar scenario was considered 
for AGN QPOs.  The most notable case is the X-ray
QPO found in RE~J1034+396 \citep{Gierlinski2008}, which had 1-hr period. 
Scaling different frequencies of QPOs in stellar-mass black holes to 
the 1-hr periodicity,
a mass range of 4$\times 10^5$--10$^7$\,$M_{\odot}$ was found for the SMBH
in RE~J1034+396. We note that this source is also a NLSy1 galaxy, the same
as J0849+5108. If we apply the same scenario to the target, an extremely
massive black hole ($>10^9\ M_{\odot}$) would be implied, in contradiction to 
the general $M_{\rm BH}$ range of NLSy1 {and the estimated mass of the SMBH
in J0849+5108. A related scenario is that the jet is precessing, caused by
the Lense-Thirring precession of the accretion disk around the black hole. This
scenario has been widely considered for QPOs seen in stellar-mass black hole
systems (see \citealt{Ingram2020} and references therein). However the above
mass-mismatch problem remains as
the timescales of the periodicities are expected to scale with $M_{\rm BH}$.}

{Generally in the coupled disk-jet systems, periodic changes in an accretion
disk would be reflected by jet activities. There are several scenarios that
could lead to the QPO phenomena. Simulations show that magnetically choked 
accretion flows (or magnetically arrested disks) could produce quasi-periodic
signals \citep{Tchekhovskoy2011,McKinney2012}. However the timescales of the
predicted signals are generally short, $\sim$1 day, not matching that of
J0849+5108. For thick disks around black holes, it has been shown that
under an external perturbation, acoustic p-mode oscillations can be excited.
The excited frequencies are related to the radial epicyclic frequency
\citep{Rubio-Herrera2015a,Rubio-Herrera2015b}.
For the 176-day period and $M_{\rm BH}\sim 10^{7.4}\,M_{\sun}$, the typical
radius can be estimated to be $\sim 360\,r_g$, where $r_g$ is the gravitational
radius. The scenario of such a disk in J0849+5108 is possible. However, 
multiple 
frequencies following the ratio of 2:3... are predicted in numerical simulations
\citep{Rubio-Herrera2015a,Rubio-Herrera2015b}, which were likely seen in
the periodic radio variability cases of the blazars NRAO~530 and 1156+295
\citep{An2013,Wang2014}. In J0849+5108, we did not see multiple frequencies. The 176-day period is similar 
to that of the QPO in the blazar J1359+4011, determined to be $\sim 120$--150 
days \citep{King2013}. The Lightman-Eardley secular 
instability \citep{Lightman1974}, applicable to the black hole systems with 
high Eddington accretion rates, was suggested to explain the QPO case of 
J1359+4011. By the same token and given the high Eddington ratio property
of the NLSy1, the instability scenario should be applicable to J0849+5108
as well.}

For year-long QPOs, their indication for the existence of a binary SMBH system 
at the center of AGN has been intriguingly discussed 
(e.g., \citealt{Valtonen2008,Ackermann2015}). The secondary SMBH may induce 
an observable periodic signal, which reflects the orbital periodicity of 
the binary. However the period of 176 day is rather too short. The intrinsic
period at the local galaxy would be $176/(1+z)\simeq 111$ days.
The period suggests
a very tight orbit (a binary separation of $\sim$0.001~pc) and a quick merging
time scale ($\sim 760$~years) due to gravitational radiation, where  
the mass ratio between the two SMBHs is assumed to be 1 (see details 
in \citealt{Sobacchi2017}).

{Finally, since the radio emission of J0849+5108 arises from its jet,
the jet activity and structure could directly lead to the QPO signal.
The source is a target in the MOJAVE survey and multiple Very Long Baseline
Array (VLBA) imaging of it during 2011--2016 are available \citep{laa2018}.
We mark the peak intensities from the observations in Figure~\ref{Fig1}
(purple circles). As can be seen, their variations are generally consistent 
with those of
the OVRO light curve. We checked the individual VLBA images. Although no
images were taken at any peak of the QPO variations, a jet structure is seen
to be nearly always present in the images extending from the core 
to 3--4 milli-arcseconds.
No evidence is seen showing the QPO variations as the results of new, emerging
radio components. }
Recently a helical structure of jets has been invoked to explain the 34.5 day
QPO observed in PKS~2247$-$131 \citep{Zhou2018}. As an emitting blob in 
a jet moves
along a helical path, our viewing angle to it changes giving rise to 
quasi-periodic flux modulation due to the Doppler beaming effect. 
Helicity is likely a natural feature in magnetically dominated 
jets \citep{Chen2020}. For the case of J0849+5108, if we assume the parameters
used in \citet{Paliya2016} for modeling the spectral energy distribution,
the bulk Lorentz factor $\Gamma = 13$, the pitch angle (between the emitting 
blob's motion and the jet's axis; assumed to be half of the opening angle) 
$\phi = 0.1$ rad, the viewing angle $\psi = 3^{\circ}$, the jet observed
at the radio 15~GHz frequency would have moved $\sim$200~pc at the local galaxy
in 21 cycles (the local cycle period would be $\sim$33 yrs; 
\citealt{ck92,Zhou2018}).
The distance of $\sim$200~pc is not unreasonably long, as the jet in the nearby
AGN M87 has been seen collimated up to a distance of $\sim$300~pc 
(e.g., \citealt{Asada2014}). {Also we note that high-frequency 
(22 and 37~GHz) imaging resolved the radio core of J0849+5108, indicating a
size of 320$\times$215\,pc \citep{Berton2018}.} Therefore we may explain 
the radio QPO of
J0849+5108 as the result of the helical motion of an emitting blob. 
Aside from the QPO modulation, the radio light curve also shows relatively 
large variations, which should be caused by the emission from the whole jet.
In addition, since the \gr\ emission 
likely arises from a site close to the central SMBH, it thus does not share any
similar variations with that at the radio frequency \citep{Max-Moerbeck2014}.

In any case, our finding of a 176 day QPO at radio in J0849+5108 presents 
another interesting case among AGN QPO phenomena. As more powerful facilities 
are available or being built, more QPO cases of different time scales at 
different frequencies will be found. QPOs of different time scales (e.g., 
intra-day, monthly, and yearly) likely
do not have the same origin, and they may not appear in the same time
at different observation frequencies.
The case in J0849+5108 proves the latter. Multi-frequency studies 
of them probably will be the key for understanding their origins, by not only
helping detect a case with high confidence, since most of them appear to be
transient, but also providing more information to establish their properties
and identify the underlying physical process among different possibilities. 

\section*{Acknowledgements}
{
{We thank anonymous referee for very helpful suggestions. We also thank
F. Xie for discussion about different black hole accretion scenarios 
and M. Gu for radio jet properties.}
This research has made use of data from the OVRO 40-m monitoring program which was supported in part by NASA grants NNX08AW31G, NNX11A043G and NNX14AQ89G, and NSF grants AST-0808050 and AST-1109911, and private funding from Caltech and the MPIfR.
This research is supported by the National Key R\&D Program of China under grant No. 2018YFA0404204, and the joint foundation of Department of Science and Technology of Yunnan Province and Yunnan University $[$2018FY001 (-003)$]$.
Z. W. acknowledges the support by the National Program on Key Research 
and Development Project (Grant No. 2016YFA0400804), 
the National Natural Science Foundation of China (11633007), the Original 
Innovation Program of the Chinese Academy of Sciences (E085021002).
}




\end{document}